\documentclass[12pt]{article}

\usepackage{newtxtext,newtxmath}

\usepackage{graphicx}

\usepackage[letterpaper,margin=1in]{geometry}

\renewenvironment{abstract}
	{\quotation}
	{\endquotation}

\date{}

\makeatletter
\renewcommand{\fnum@figure}{\textbf{Figure \thefigure}}
\renewcommand{\fnum@table}{\textbf{Table \thetable}}
\makeatother

\usepackage{scicite}

\usepackage{url}

\def\scititle{
	Topological Antenna; A Non-Classical Beam-Steering  Micro-Antenna  Based on Spin Injection from a  Topological Insulator 
}
\title{\bfseries \boldmath \scititle}

\author{
	Raisa~Fabiha$^{1}$,
    Patrick~J~Taylor$^{2}$, 
	Supriyo~Bandyopadhyay$^1 \ast$\and
	\small$^{1}$Department of Electrical and Computer Engineering, Virginia Commonwealth University,\and \small Richmond, VA 23284, USA \\
	\small$^{2}$DEVCOM Army Research Laboratory, Adelphi, MD 20783, USA\and
	\small$^\ast$Corresponding author. Email: sbandy@vcu.edu
}

\begin{document} 

\maketitle
\vspace{-0.4in}
\begin{abstract} \bfseries \boldmath
Antennas are the quintessential means to communicate information over long distances via electromagnetic waves. Traditional antennas have two shortcomings that have prevented miniaturization: (1) their radiation efficiencies plummet and (2) they radiate  isotropically when miniaturized to small fractions of the radiated wavelength. Here, we report a new genre of non-classical antennas that overcome these limitations by employing non-traditional principles and harnessing topological insulators.  An {\it alternating} charge current of frequency 1-10 GHz injected into a thin film of a three-dimensional topological insulator (3D-TI) injects a spin current of alternating spin polarization into a periodic array of cobalt nanomagnets deposited on the surface of the 3D-TI. This generates spin waves in the nanomagnets, which radiate electromagnetic waves in space, thereby implementing an antenna. The frequency of the electromagnetic wave is the same as that of the current. The antenna dimension is only 0.6-1.8\% of the free space wavelength and yet it radiates with an efficiency several orders of magnitude larger than the theoretical limit for conventional antennas. Furthermore, it radiates anisotropically (despite being a ``point source'') and one can change the anisotropic radiation pattern by changing the direction of the injected alternating charge current, which changes the spin wave patterns within the nanomagnets because of {\it spin-momentum locking} in the 3D-TI. This enables beam steering without the use of a phased array. We have overcome several limitations of classical antennas by harnessing the quantum mechanical attributes of a quantum material, namely a 3D-TI.
\end{abstract}

\section{Introduction}

Three-dimensional topological insulators (3D-TI) are remarkable quantum materials exhibiting spin-momentum locking of the surface states and spin-polarized surface current \cite{moore,li,kang}. The latter has been harnessed to inject spin currents into ferromagnets deposited on the surface of a 3D-TI via charge current injection, resulting in the switching of magnetization at room temperature \cite{mellnik, wang}.  The conversion ratio, which is the ratio of the spin current density to the charge current density can significantly exceed unity \cite{mellnik}, making this a very efficient mechanism of converting a charge current into a spin current.

\begin{figure}[hbt!]
\centering
\includegraphics[width=0.91\textwidth]{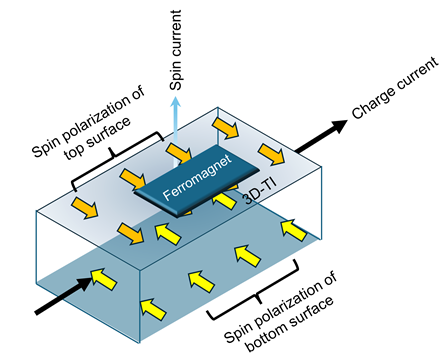}
\caption{\small A ferromagnet deposited on a three-dimensional topological insulator. The directional relationships between the charge current, injected spin current and spin polarization of the injected spin current (which is same as that of the top surface)  are shown. The magnetization of the ferromagnet will be aligned along the spin polarization. An alternating charge current will result in oscillating spins.}
\label{fig:currents}
\end{figure}

\begin{figure*}[hbt!]
\centering
\includegraphics[width=0.91\textwidth]{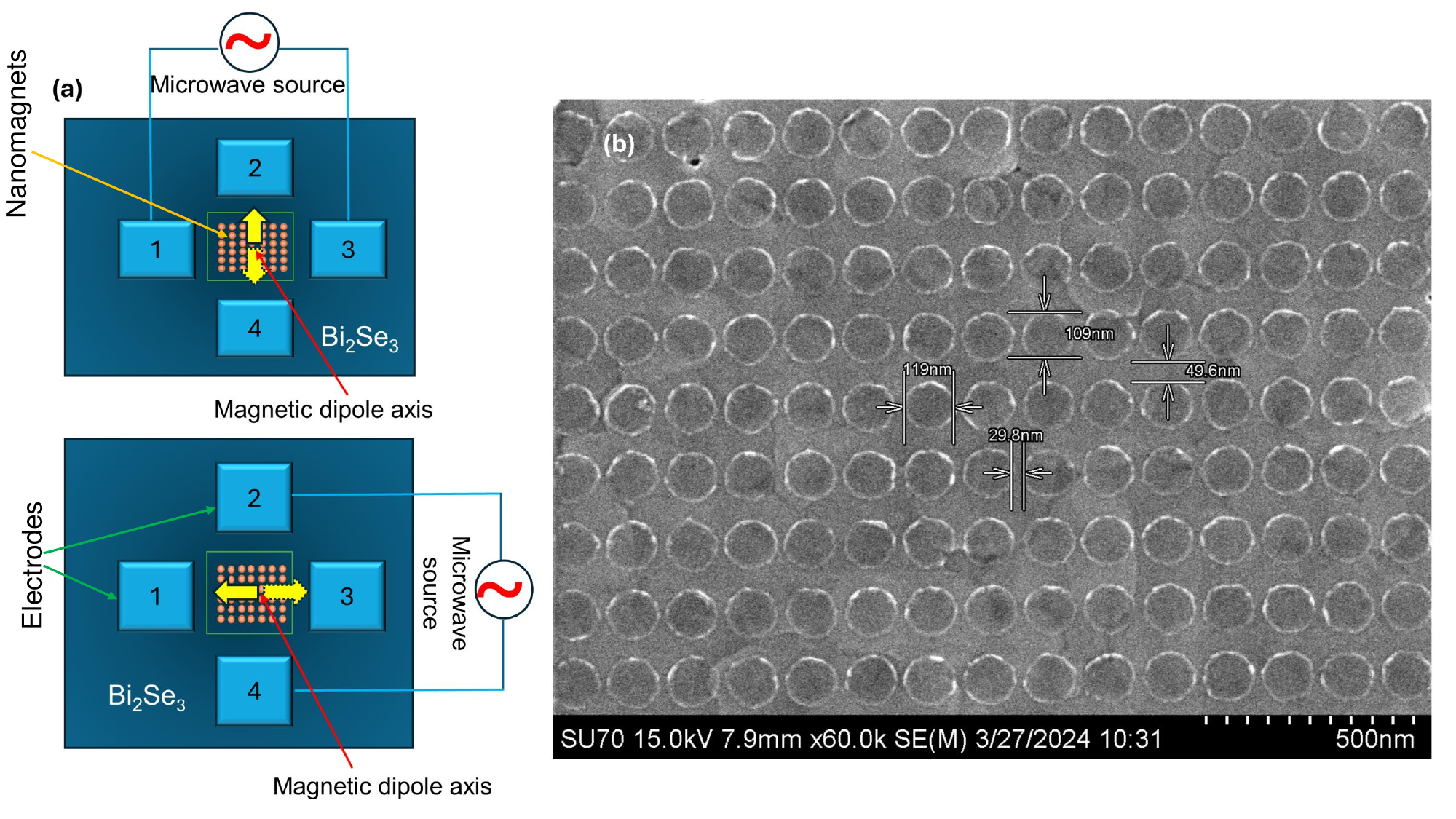}
\caption{\small (a) Schematic of the nano-antenna and two different ways of passing an alternating charge current between antipodal electrode pairs leading to two different directions of charge current flow and hence two different axes of spin polarization in the spin current injected into the nanomagnets. The axes of the oscillating spins are shown for the two cases. The top configuration is referred to as ``orientation 1'', and the bottom as ``orientation 2''. (b) Scanning electron micrograph of the fabricated nanomagnet array. There are 15.36 million nanomagnets in the array, making the total antenna area about 0.003 cm$^2$.}
\label{fig:TI-antenna}
\end{figure*}


When a charge current is injected into a 3D-TI, the two opposite surfaces become oppositely spin polarized \cite{mellnik} as shown in Fig. \ref{fig:currents}. This results in a spin current flowing perpendicular to the surfaces that can inject spins into a nanomagnet placed on the top surface. The spin current flows perpendicular to the charge current and its spin {\it polarization} is perpendicular to both the spin current and the charge current, as shown in Fig. \ref{fig:currents}. If the direction of the charge current flow is reversed, then the spin polarization of the spin current should be reversed as well because of the spin-momentum locking. Therefore, if an {\it alternating} charge current is injected into a 3D-TI, it will inject a spin current of {\it alternating} spin polarization into a nanomagnet placed on the surface. This will exert an alternating spin-orbit torque (SOT) on the nanomagnets \cite{wang},  causing spin waves to be generated within them \cite{morrison}. These spin waves are confined (standing) spin waves within the nanomagnets, which can transfer energy to electromagnetic waves that radiate into the surrounding medium \cite{raisa, kuo} with the frequency of the charge current. This implements a novel ``antenna''.  

Here, the spin waves transfer some of their energies into electromagnetic waves, thus producing the radiation and antenna action. The efficiency of energy transfer (and hence the radiation efficiency of the antenna) depends on the strength of magnon-photon coupling, which can be strong in the presence of interface spin-orbit torque (SOT) \cite{salikhov}.

Section 3 of the Supplementary Material provides a classical phenomenological proof of how  the alternating spin-orbit torque resulting from spin injection from the topological insulator into the nanomagnets will result in electromagnetic emission. This is the theory behind the antenna presented here and it is based on coupled Landau-Lifshitz-Gilbert equation and Maxwell's equation.

We can introduce an additional twist in this phenomenon. Because of the fixed relationship between the axis of the oscillating spin polarization in the nanomagnets and the direction of the charge current density (due to spin-momentum locking) [see Fig. \ref{fig:TI-antenna}], we can rotate the axis of the oscillating spins by rotating the direction of the alternating current flow. This will then rotate the electromagnetic {\it radiation pattern} [see Section 3 of the Supplementary Material for the proof], thereby implementing ``beam steering''. 

In this work, we have demonstrated both the antenna functionality and the beam steering capability. Normally, beam steering requires a phased array consisting of multiple radiating elements. The phased array is several times larger than the radiated free-space wavelength. Here, we have demonstrated the possibility of beam steering with a {\it single} element that is roughly two orders of magnitude {\it smaller} than the radiated free-space wavelength.

\subsection{Design of the micro-antenna}

The schematic of our micro-antenna is shown in Fig. \ref{fig:TI-antenna}(a). It consists of a periodic array of $\sim$15 million slightly elliptical Co nanomagnets delineated on a 20-nm thick Bi$_2$Se$_3$ film (a 3D-TI) deposited on a (001) PMN-PT substrate. Four contact pads surround the nanomagnet array and are used for charge current injection in two mutually perpendicular directions.  The substrate is piezoelectric but in this work the piezoelectricity plays no role and can be ignored. The major axis of a nanomagnet is $\sim$120 nm, the minor axis is $\sim$110 nm and the thickness is 6 nm with a 5 nm Ti adhesion layer underneath each nanomagnet. 
The edge-to-edge separation between nearest neighbors along the major axis is $\sim$30 nm and  along the minor axis is $\sim$50 nm. Fig. \ref{fig:TI-antenna}(b) shows a scanning electron micrograph of the nanomagnet array with all relevant dimensions.

In a 20 nm thick TI film, there will very likely be some bulk conduction in addition to conduction via topological surface states which are spin-momentum locked. Separating bulk conduction from surface conduction is very difficult and complicated. We have investigated this for (Bi,Sb)$_2$(Te,Se)$_3$ tetradymite materials in the past \cite{Taylor}, but not for Bi$_2$Se$_3$. This is outside the scope of this work. However, the antenna {\it functionality} does not depend on what fraction of the conduction is through the bulk and what fraction is through the surface. As long as there is {\it some} surface conduction, no matter what fraction of the total conduction it is, the effect exists. If one can increase the ratio of surface to bulk conduction, that will obviously increase the radiation efficiency since the bulk component does not  contribute to radiation, but any {\it non-zero} surface conduction will produce the electromagnetic radiation. Hence, differentiating between surface and bulk conduction is not an important requirement at this time.

We point out that because we have isolated Co nanomagnets and not a continuous thin film of Co, the so-called ``current shunting'' problem \cite{liu}, which reduces the efficacy of spin injection from a topological insulator (TI) into a ferromagnet, is greatly mitigated. There is simply no current shunting path that is long enough to divert much of the spin polarized current away from the surface of the TI. The use of nanomagnets, as opposed to a thin film, also suppresses the formation of  eddy current loops in the ferromagnets and this reduces eddy current loss, thereby increasing the antenna's radiation efficiency.  

\section{Results and Discussion}

\subsection{Antenna activation and beam steering}

\begin{figure}[!b]
\centering
\includegraphics[width=0.91\textwidth]{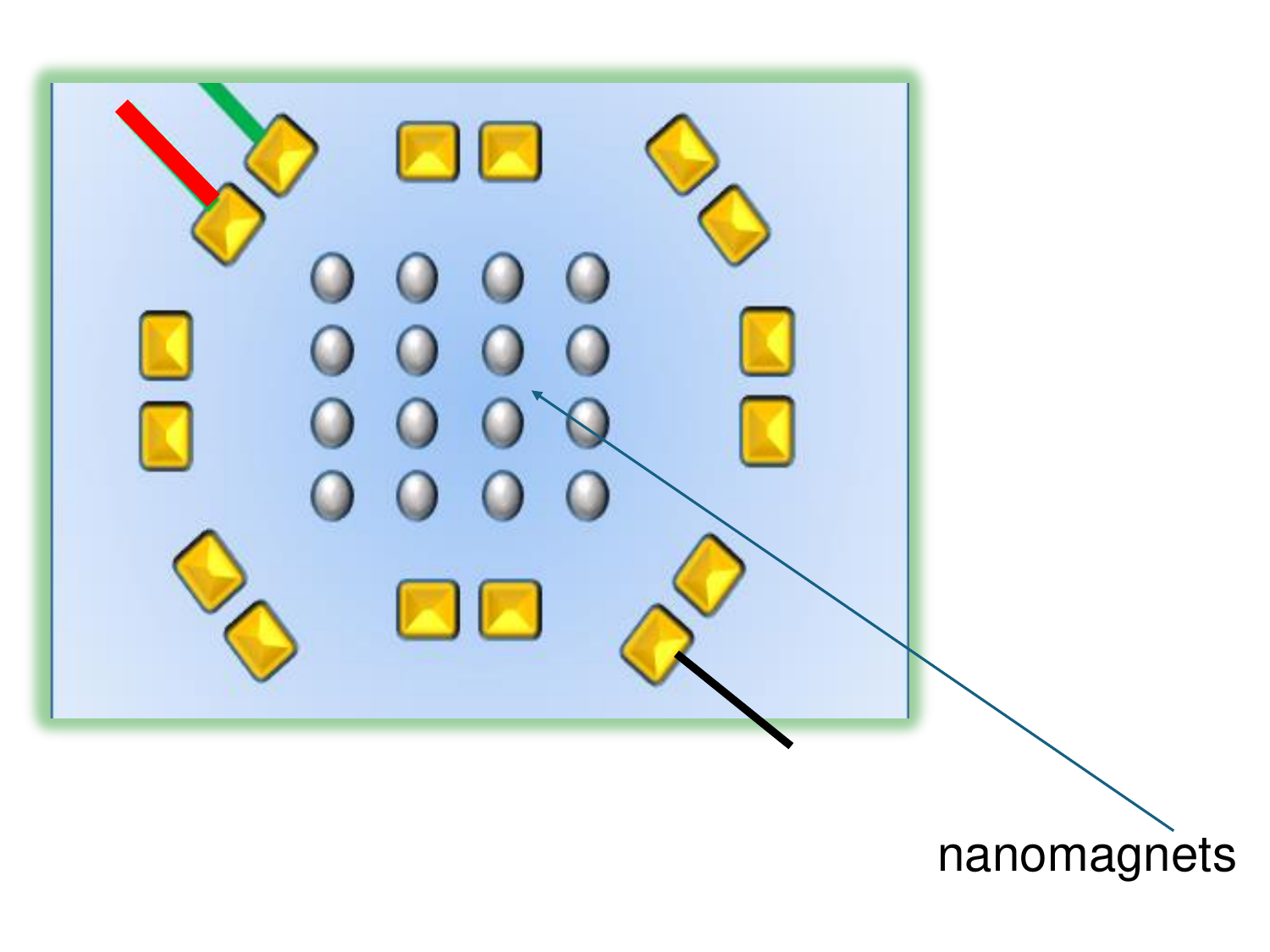}
\caption{\small Modality of beam steering. By injecting current between one fixed electrode (connected to a black line)  and all the others sequentially using a multiphase clock, we can inject current sequentially in different directions to make the principal lobe of the radiation pattern scan 360$^{\circ}$ and thus implement beam steering.}
\label{fig:array}
\end{figure}

An alternating charge current of 1- 10 GHz frequency  is injected between either electrode pair 1 and 3, or electrode pair 2 and 4 in Fig. \ref{fig:TI-antenna}, to inject spin currents of alternating spin polarization  into the nanomagnets.
The spin oscillation axis for either electrode pair activation is shown in Fig. \ref{fig:TI-antenna}(a).   Changing this axis by changing the direction of the charge current flow will change the spin wave pattern within the nanomagnets, i.e., the $x$-, $y$- and $z$-components of the oscillating magnetic field associated with the spin wave (both amplitude and phase). This oscillating magnetic field and the associated oscillating electric field determine the Poynting vector of the emitted electromagnetic radiation and therefore determine the radiation pattern, as explained in Section 5 of the Supplementary Material. The radiation patterns in different planes will therefore depend on which of the two electrode pairs -- (1,3) or (2,4) -- in Fig. \ref{fig:TI-antenna} is activated, i.e., between which two antipodal electrodes the charge current flows. This leads to the beam steering capability; by activating either of the two antipodal pairs at the exclusion of the other, we can steer the principal lobe of the  radiated beam to two different directions. By extension, if we place more electrodes around the nanomagnet array as shown in Fig. \ref{fig:array}, then by injecting currents between different electrodes pairwise using a multi-phase clock, we can continuously steer or scan the beam in space in the manner of an {\it active electronically scanned array (AESA)}. Normally, beam steering or active scanning requires a phased array antenna whose linear dimensions are several times larger than the radiated free-space wavelength, whereas here we are able to do that with an antenna whose linear dimensions ($\sim$550 $\mu$m) are 2-3 orders of magnitude smaller than the free-space wavelength at the frequencies we tested at (1-10 GHz; free space wavelength 3-30 cm). This would lead to an ultra-miniaturized beam scanner.


\subsection{Electromagnetic spectrum}

\begin{figure*}[htbp]
\centering
\includegraphics[width=0.8\textwidth]{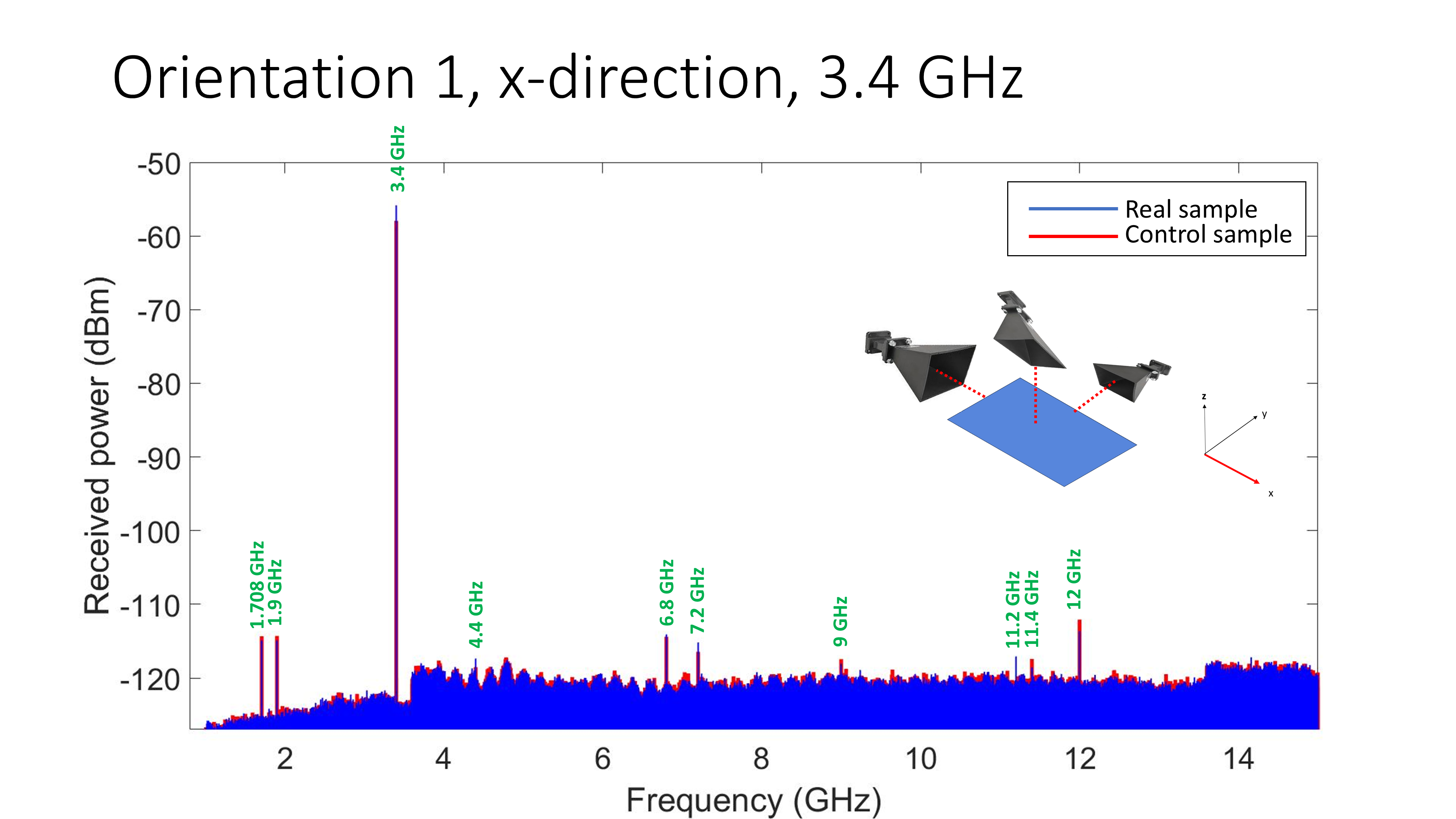}
\includegraphics[width=0.8\textwidth]{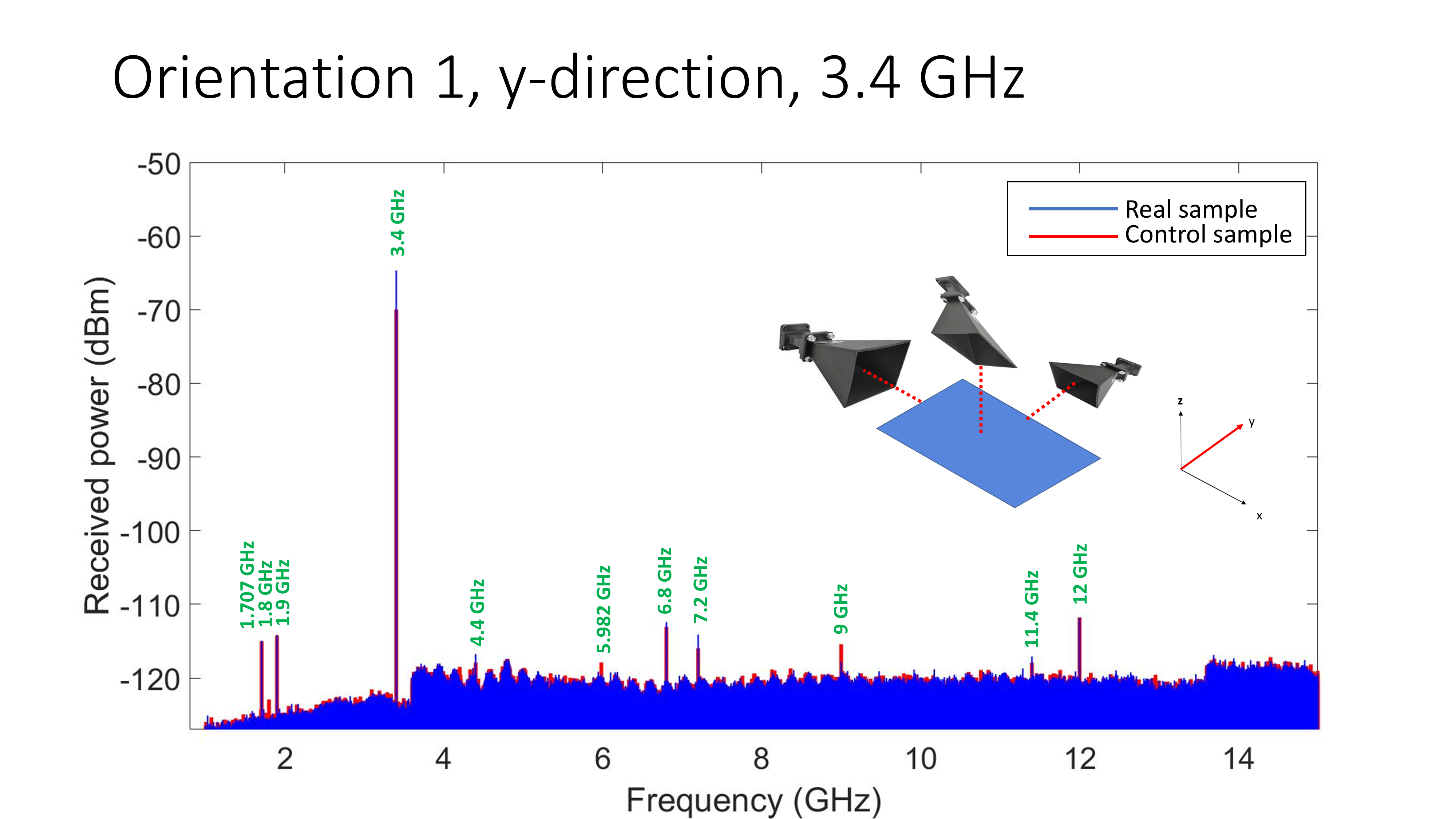}
\caption{\small Spectra of the electromagnetic emission when the alternating current frequency is 3.4 GHz. The input power from the alternating current source is 31 mW (15 dbm). The spectra are shown for two different placements of the horn antenna with respect to the sample. The horn was placed in the plane of the nanomagnets facing the two different edges. The line joining the horn and the sample is along the direction $x$ and along the direction $y$ in the two cases. In both cases, current was passed between contact pads 1 and 3.}
\label{fig:spectrum}
\end{figure*}

\begin{figure*}[!b]
\centering
\includegraphics[width=0.91\textwidth]{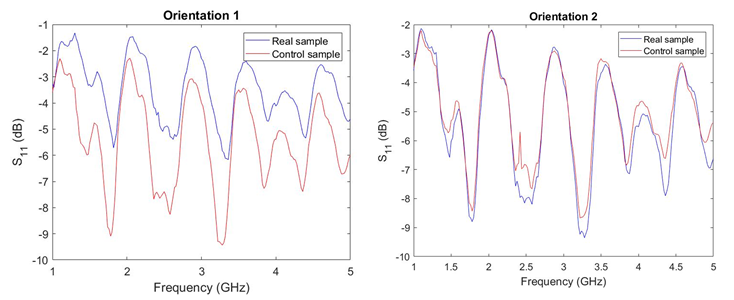}
\caption{\small Spectrum of the scattering parameter S$_{11}$ of the real and the control sample when current is injected in two mutually orthogonal directions labeled as orientations 1 and 2.}
\label{fig:S11}
\end{figure*}

We measured the spectra of the electromagnetic radiation emitted by the samples in an AMS-5701 anechoic chamber with a noise floor of -120 dbm. All measurements are done at room temperature. The detector was a polarization-sensitive horn antenna and a spectrum analyzer was connected to it to measure the spectrum of the received emission. The sample was placed at a distance of 81 cm from the horn antenna to ensure that we are always measuring the far-field radiation at the excitation frequency (3-10 GHz). The input power from the alternating current source was set to 15 dbm (31 mW). In Fig. \ref{fig:spectrum}, we show the spectra for two different placements of the detecting horn antenna with respect to the sample [see the inset of the figure for the horn antenna placement], when the alternating charge current frequency is 3.4 GHz. These two spectra are very similar confirming the expectation that the {\it spectrum} (i.e., the frequency components of the emitted radiation) should have no directional dependence.

The electromagnetic radiation detected by the horn antenna is of course not solely due to emission from the nanomagnets, but also has contributions from  extraneous sources of radiation, such as the contact pads and wiring. In an attempt to separate out the extraneous contributions, we fabricated two sets of samples that are otherwise nominally identical, except one has the nanomagnets and the other does not. We call the latter ``control sample'' and measure the spectrum of the radiation it emits for comparison with that of the real sample which contains the nanomagnets. In Fig. \ref{fig:spectrum} we show the measured spectra for both the real and the control sample. At 3.4 GHz, the power received from the real sample is a few dBm larger than that from the control sample, which hints at the fact that the nanomagnets are radiating electromagnetic waves.

Unfortunately, we cannot just subtract the power received from the control sample from that received from the real sample to obtain the power radiated by the nanomagnets. This is because of wave {\it interference}. The waves emitted by the nanomagnets and those emitted by the peripherals (such as the contact pads) interfere at any point in space so that the power received from the real sample is {\it not} the arithmetic sum of the power received from the nanomagnets and from the peripherals. Consequently, the difference between the power received from the real and control samples at any point in space is not due to the power emitted by the nanomagnets. Furthermore, in our case, the difference is small. When the line joining the detecting horn antenna and the sample is along the $x$-direction [Fig. \ref{fig:spectrum} top panel], the power detected by the horn antenna from the real sample is -55 dbm = 3.16 nW while that detected from the control sample is -58 dbm = 1.6 nW.
The difference of 1.5 nW, albeit well above the noise floor in the anechoic chamber [which was -120 dbm (1 pW)], is too small to draw any quantitative conclusions. Moreover, we cannot determine how much of this difference actually accrues from the nanomagnets and how much is due to unavoidable slight differences between the peripherals (i.e. contact pads, etc.) in the real sample and the control sample.  Hence, we cannot (accurately) measure the radiation from the nanomagnets alone and that is unavoidable. 

When the line joining the horn and the sample is along the $y$-direction [Fig. \ref{fig:spectrum} bottom panel], the power detected from the real sample is -65 dbm (0.3 nW) and that from the control sample is -70 dbm (0.1 nW). Again, the difference between the two is too small to draw any inference about the magnitude of power radiated by the  nanomagnets. However, we can convincingly claim that the nanomagnets are radiating {\it some power} because the radiation {\it patterns} of the real and the control sample are {\it markedly different (by 3 orders of magnitude) in some directions}, as we show later. This could not happen without the nanomagnets radiating electromagnetic waves. It is just that the uncertainties arising from wave interference and unintentional differences between the peripherals in the real and control samples prevent us from accurately measuring the power emitted by the nanomagnets alone. As a result, we also cannot measure the radiation efficiency of our micro-antenna accurately, although antennas of this genre are known to radiate very efficiently \cite{raisa,carman,justine,saibal}, far more efficiently than conventional antennas that are constrained by the Harrington limit \cite{Harrington,skrivervik}.

Note that the dominant peak in the spectrum is at the frequency of the alternating current injected into the 3D-TI as expected because that is the frequency of the oscillating spins. There are also satellite peaks in the spectra of both the real sample and the control sample, usually at the same frequencies in both samples. They are much weaker than the main peak, always below -110 dbm or 10 pW, and almost surely accrue from extraneous sources. Unfortunately, we do not have enough spectral resolution to measure the width of the peaks, but the widths are obviously very narrow, resulting in high quality factor. 

We have also measured the spectrum of the scattering parameter $S_{11}$ with a vector network analyzer and that data are shown in Fig. \ref{fig:S11} for both the real and the control sample. We show the results for two mutually orthogonal directions of the alternating charge current injection (`orientation 1' and `orientation 2'). Note that the $S_{11}$ parameter for the real sample tracks that of the control sample fairly closely, but is usually (not always) equal to or greater than that of the control sample, indicating that the nanomagnets are causing some additional reflection of the input signal.

The reflection is of course unwanted, but it is often unavoidable since most antennas are reactive and hence not impedance matched to a standard 50-ohm microwave source. Consequently, the $S_{11}$ parameter is often high, as is the case here. However, this is ultimately not an issue since antennas are usually impedance matched to the source with a low insertion loss impedance matching transmission line which reduces $S_{11}$ significantly \cite{abracon}. We did not employ any impedance matching network here since that is outside the scope of this work. We are not concerned with antenna optimization, but rather the new physics which has nothing to do with impedance mismatch.

\subsection{Radiation patterns} 

We measured the radiation patterns of the micro-antenna in three different planes -- the plane of the nanomagnets and the two transverse planes.  The patterns were measured at frequencies of 3.4, 5 and 10 GHz (free space wavelengths of 3, 6 and 9 cm) in an AMS-8100 anechoic chamber. The input power was 1 mW.

The radiation detector is again a polarization-sensitive horn antenna that is always placed at a distance of 284.5 cm from the sample, which means that we are always detecting the far-field radiation pattern.

\begin{figure*}[!hbt]
\centering
\includegraphics[width=0.91\textwidth]{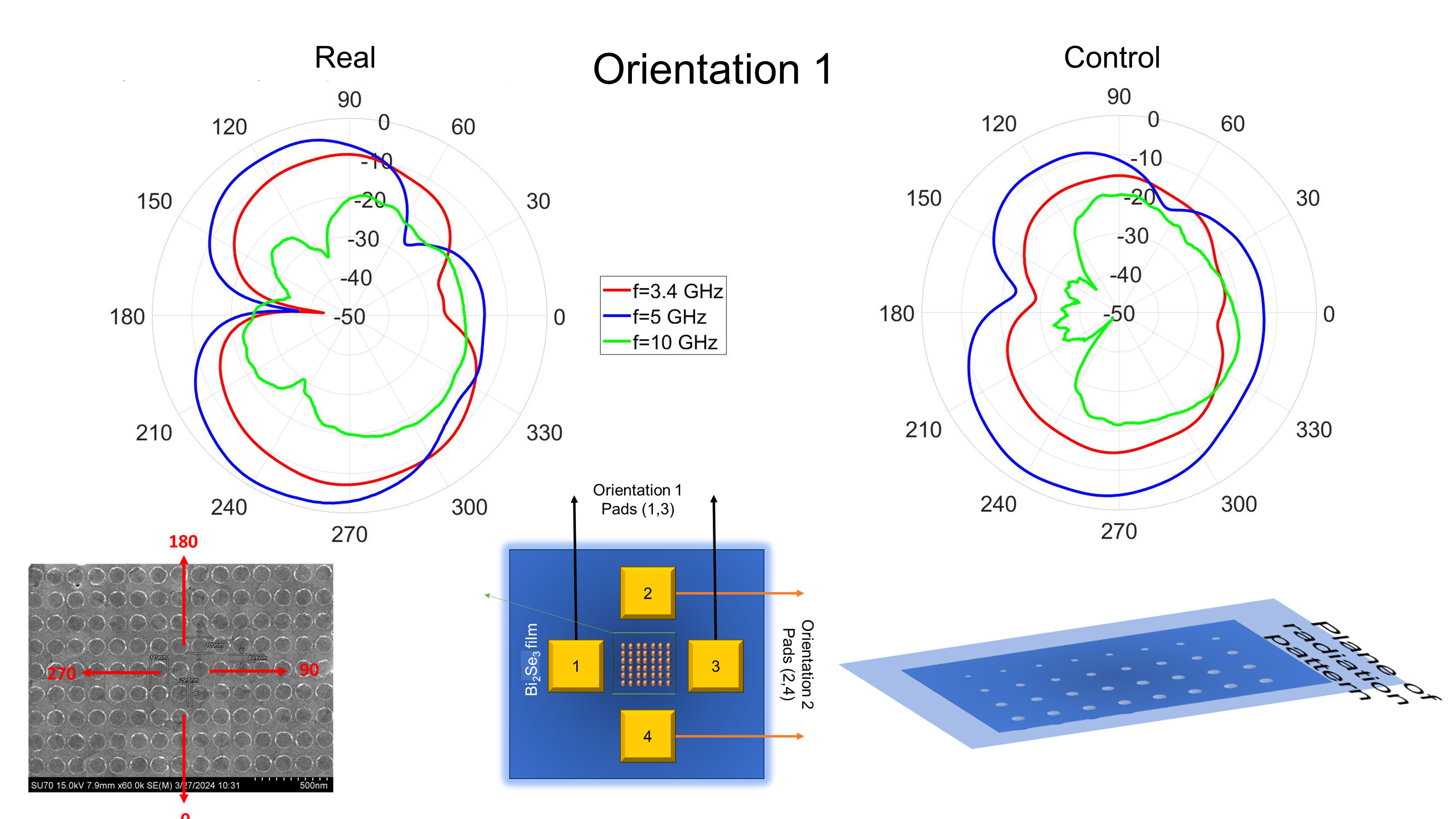}
\vspace{1.0cm}
\includegraphics[width=0.91\textwidth]{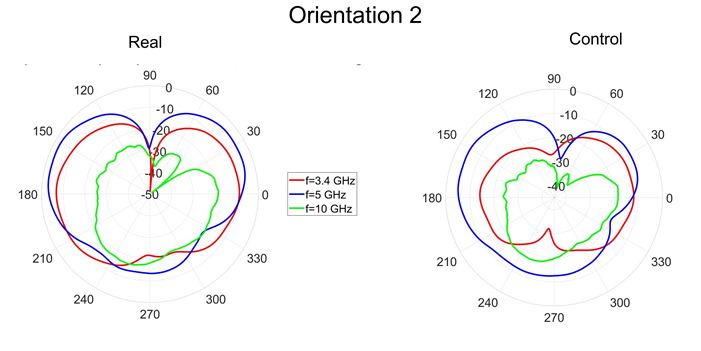}
\caption{\small The radiation pattern for the horizontal polarization of the electromagnetic wave detected by the detector at three different frequencies (3.4, 5 and 10 GHz) in the plane of the nanomagnets. The patterns are shown for both the real sample and the control sample for two different directions of charge current flow (orientation 1 and orientation 2). The patterns are plotted in dbi. The difference between the radiation patterns of the real and control sample is small but large enough to establish that the nanomagnets are radiating.}
\label{fig:control}
\end{figure*}

In Fig. \ref{fig:control}, we show the radiation patterns in the plane of the nanomagnets for both the real and the control sample, for horizontal polarization of the emitted radiation, and for both directions of alternating charge current injection (orientations 1 and 2). In some directions in the plane of the nanomagnets, the received power from the control sample is {\it more} than that from the real sample. This apparent anomaly can be explained by destructive interference. The waves transmitted by the peripherals (contact pads, etc.) and those transmitted by the nanomagnets interfere at any point in space. In some directions, the interference at the location of the horn antenna is {\it destructive}, which makes the power received from the real sample less than that from the control sample. 

However, in some other directions, the power received from the real sample {\it far exceeds} that from the control sample, which is a direct confirmation that the nanomagnets are radiating. For example, at 10 GHz and orientation 1 [see Fig. \ref{fig:control}], the measured gain in the 225$^{\circ}$ direction [horizontal polarization] is about -50 dbi for the control sample and -20 dbi for the real sample. This difference of 30 dbi (a factor of 1000$\times$) is large enough to claim that the nanomagnets are definitely radiating, at least in the 225$^{\circ}$ direction at 10 GHz, and perhaps in other directions as well. This confirms that the nanomagnets radiate electromagnetic waves, which vindicates the antenna operation.

The radiation from the nanomagnets is, however, extremely directional and far from isotropic because the difference between the radiation from the real sample and the control sample is strongly direction-dependent. This means that the nanomagnets are radiating {\it anisotropically}. On the face of it, this anisotropy is surprising since the lateral dimension of the entire nanomagnet array is 2-3 orders of magnitude smaller than the free-space electromagnetic wavelength at all measurement frequencies, which should make the antenna effectively a ``point source'' that should radiate isotropically. Yet, it does not. Why this happens is explained in Section 5 of the Supplementary Material. This section also explains why it would be very difficult to reproduce the radiation pattern from one sample to another because that would require a degree of fabrication perfection not currently available anywhere. We will have to reproduce the size, shape, surface roughness, thickness, etc. of every one of the 15 million nanomagnets since they determine the spin wave patterns, which, in turn, determine the radiation pattern. This would be impossible to achieve with any state-of-the-art nanofabrication technology.

\begin{figure*}[!hbt]
\centering
\includegraphics[width=0.8\textwidth]{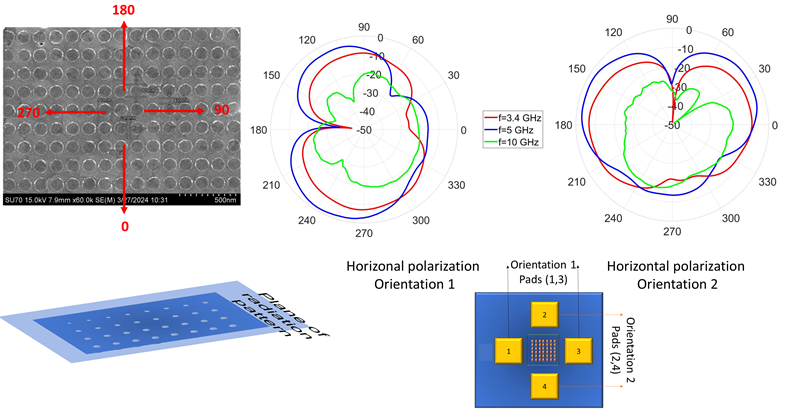}
\vspace{1cm}
\includegraphics[width=0.8\textwidth]{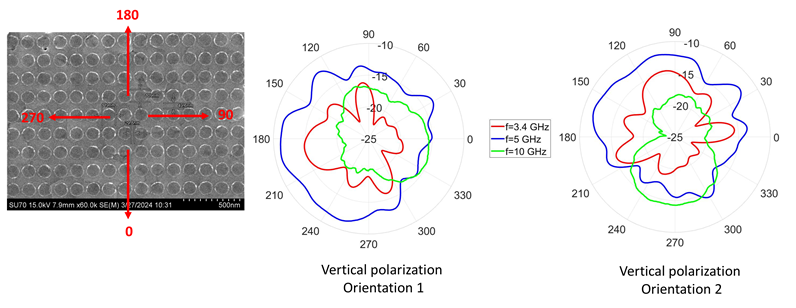}
\caption{\small The radiation pattern of the real sample at three  different frequencies (3.4, 5 and 10 GHz) in the plane of the nanomagnets. The patterns are shown for both horizontal and vertical polarizations of the emitted beam. The radiation pattern for the control sample was already shown in Fig. \ref{fig:control} for horizontal polarization. For vertical polarization, there are similar differences between the radiation patterns of the real and control samples and hence the radiation pattern for the control sample is not shown. The horizontally polarized component is much stronger than the vertically polarized component in this plane. Note the very significant differences in the radiation patterns between orientation 1 (charge current flows between contact pads 1 and 3) and orientation 2 (charge current flows between contact pads 2 and 4) for both polarizations. This difference allows one to change the radiation pattern significantly by changing the direction of current flow, attesting to the beam steering capability. Note that for the two lower frequencies of 3.4 GHz and 5 GHz, {\it the radiation pattern rotates by approximately 90$^{\circ}$ when we rotate the direction of current flow by 90$^{\circ}$}. This is a direct manifestation of beam steering. }
\label{fig:pattern1}
\end{figure*}

Perhaps the most interesting observation is that the radiation pattern depends strongly on the direction of charge current flow (note the differences between orientations 1 and 2 in Fig. \ref{fig:pattern1}). This is not surprising since the charge current's direction determines the axis of the alternating spin polarization within a nanomagnet owing to {\it spin-momentum locking}. Thus, by changing the direction of current flow, we can change the spin polarization axis and hence  the spin wave pattern  in different nanomagnets. Since it is the spin waves that radiate the electromagnetic waves, changing the spin wave patterns will change the radiation pattern, which allows beam steering. 

One will also notice that changing the direction of the current not only changes the radiation pattern of the real sample, but it also changes that of the control sample! On the face of it, this may seem to contradict our hypothesis, but this is actually expected in our configuration. The reason for this is explained in Section 6 of the Supplementary Material. Because of this feature, we have to check if the radiation patterns of the real sample and the control sample {\it change by approximately the same amount when we change the current direction}. If they do, then we must conclude that the radiation pattern of the nanomagnets does {\it not} depend on current direction.  But that does not happen. They change in different ways, especially at the lowest frequency of 3.4 GHz, which tells us that the radiation pattern of the nanomagnets changes if we change the current direction. 

\clearpage

\begin{figure*}[!hbt]
\centering
\includegraphics[width=0.8\textwidth]{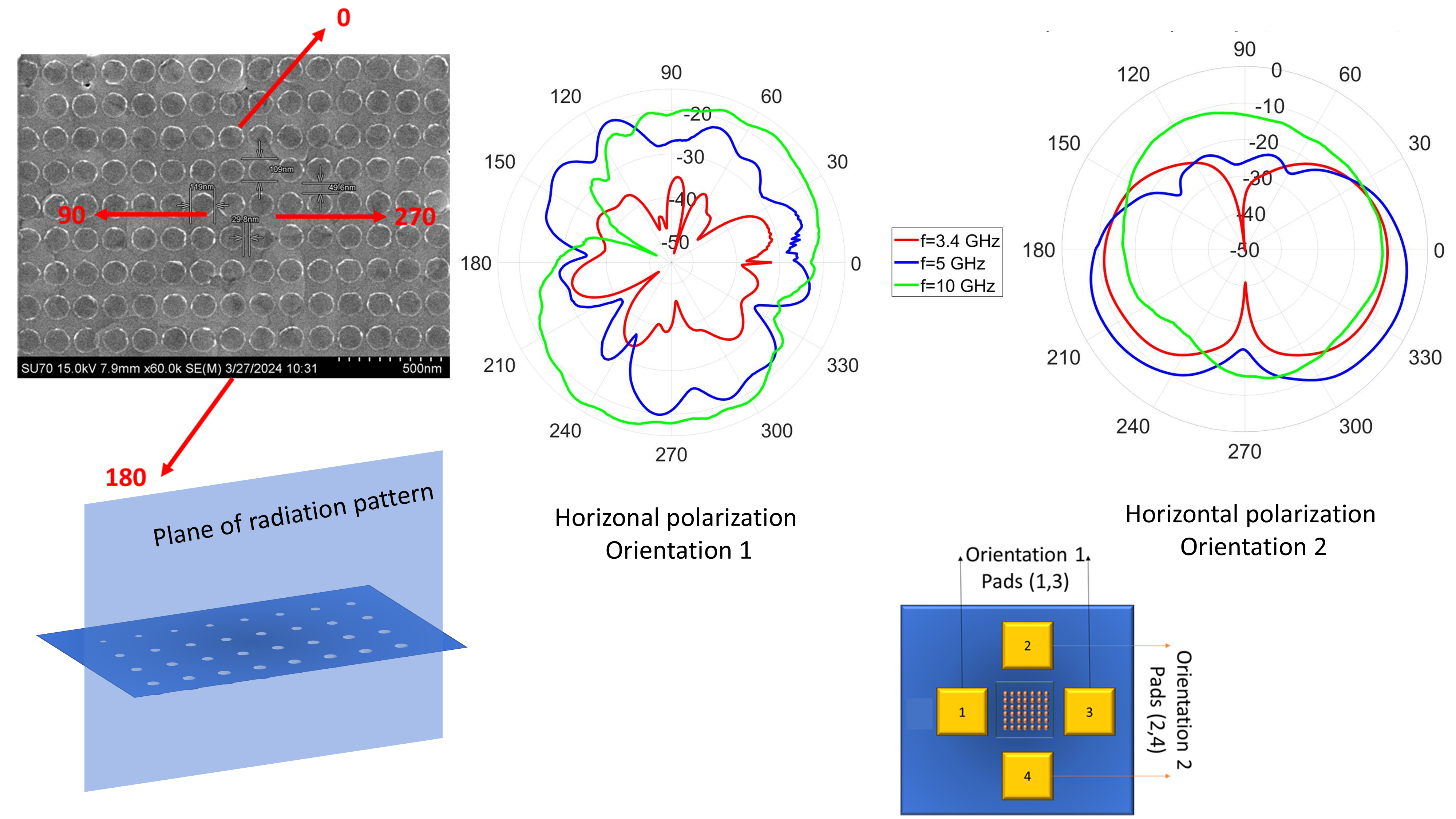}
\includegraphics[width=0.8\textwidth]{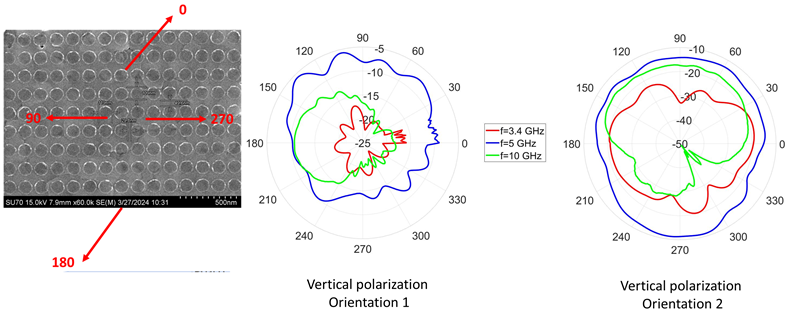}
\caption{\small The radiation pattern of the real sample at three different frequencies of 3.4, 5 and 10 GHz in a plane transverse to that of the nanomagnets. The radiation pattern of the control sample is again quite different from that of the real sample in some directions (as in Fig. \ref{fig:pattern1}) and is not shown here to avoid repetition. The patterns are shown for both horizontal and vertical polarizations.  Again, note the significant differences in the radiation patterns for orientation 1 (charge current flows between contact pads 1 and 3) and orientation 2 (charge current flows between contact pads 2 and 4).}
\label{fig:pattern2}
\end{figure*}

\clearpage

\begin{figure*}[!hbt]
\centering
\includegraphics[width=0.8\textwidth]{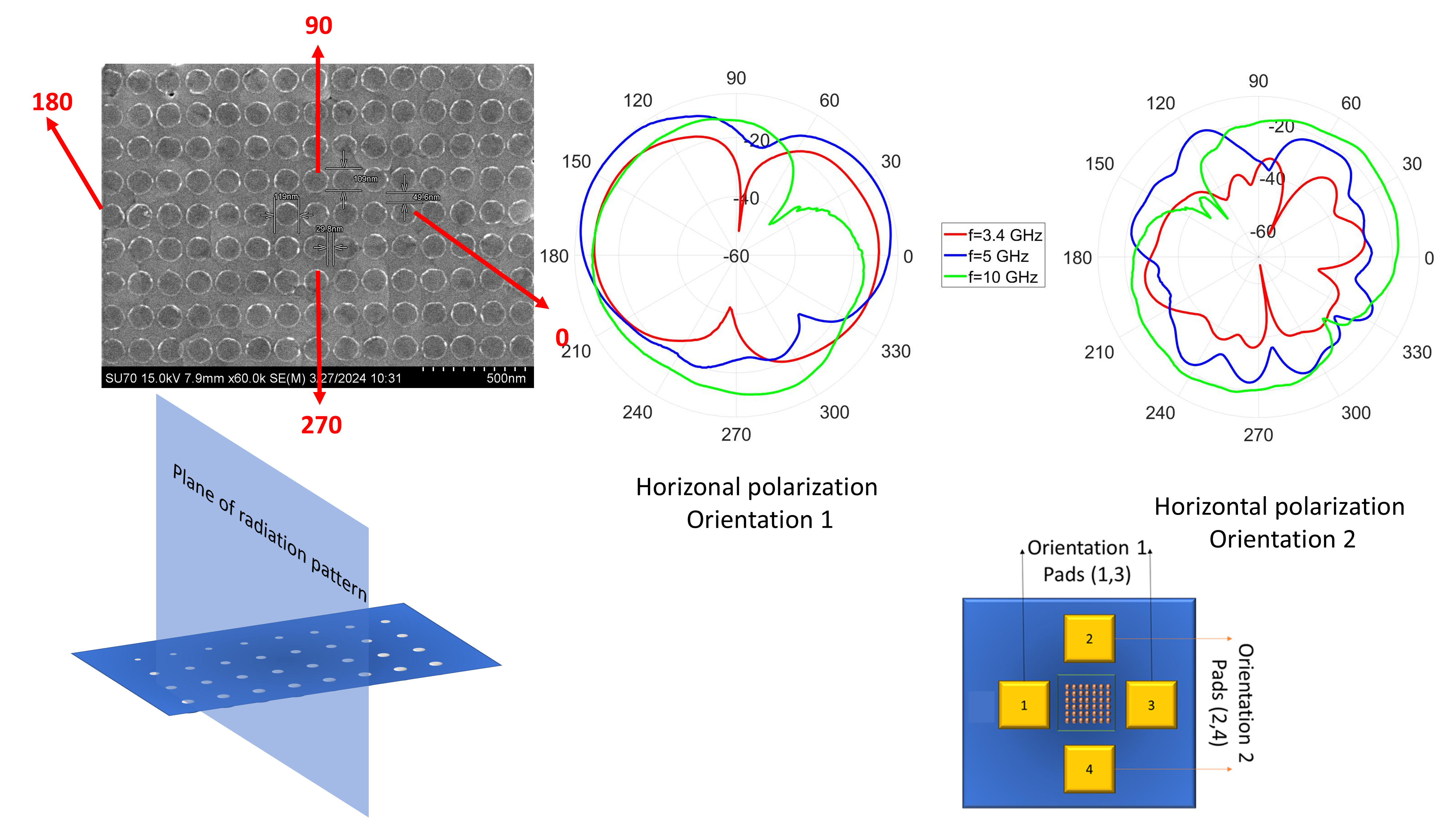}
\vspace{1cm}
\includegraphics[width=0.8\textwidth]{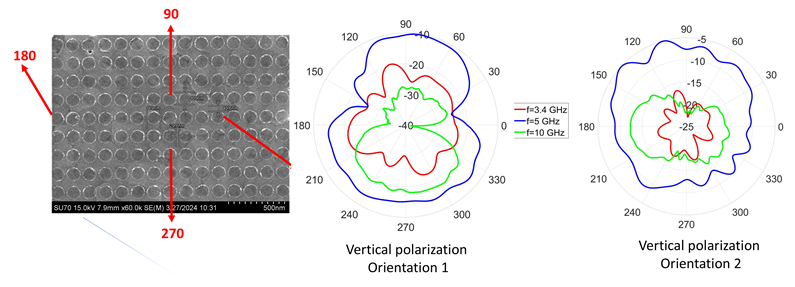}
\caption{\small The radiation pattern of the real sample at frequencies of 3.4, 5 and 10 GHz in the other plane transverse to that of the nanomagnets. Again, the radiation pattern of the control sample is quite different from that of the real sample in some directions (as in Fig. \ref{fig:pattern1}) and is not shown here to avoid repetition. The patterns are shown for both horizontal and vertical polarizations. Again, note the stark differences in the radiation patterns for orientation 1 (charge current flows between contact pads 1 and 3) and orientation 2 (charge current flows between contact pads 2 and 4).}
\label{fig:pattern3}
\end{figure*}

\clearpage

In Fig. \ref{fig:pattern1} we note that at the two lower frequencies of 3.4 and 5 GHz, rotating the direction of the current by 90$^{\circ}$ {\it also rotates the radiation pattern approximately by 90$^{\circ}$ for the horizontal polarization}, which is the dominant polarization in the emitted beam.  This feature is not observed in the control sample in Fig. \ref{fig:control} which tells us that it originates from the nanomagnets. We also do not see this feature so prominently at the highest frequency of 10 GHz in the real sample because at that high frequency, the spin oscillation may not be able to keep up in synchrony with the charge oscillation, thereby weakening beam steering. The lower frequency feature, however, is a direct manifestation of ``beam steering'' which can only happen if the nanomagnets are radiating. {\it It is a unique modality of beam steering not possible with conventional antennas and made possible here by the unique spin-momentum locking property of a topological insulator. Here, we are steering the beam using an extreme sub-wavelength single antenna, whereas the conventional modality would require a phased array that consists of multiple antennas, each larger than the wavelength}.

This beam steering capability also attests to the fact that there must be a good degree of surface conduction in the TI film since spin-momentum locking is a property of surface states and not bulk states. However, disentangling surface conduction from bulk conduction to determine their relative contributions is neither within the scope of this work, nor does it affect the antenna operation or principle in any way. 

The beam steering effect is less pronounced in the radiation patterns in the two planes that are transverse to the plane of the nanomagnets, as can be seen in Figs. \ref{fig:pattern2} and \ref{fig:pattern3}. However the radiation intensities in these two planes are weaker than in the plane of the nanomagnets, which is not surprising since the spin oscillation takes place in the plane of the nanomagnets.

Finally, we wish to address an obvious question, namely what would have been the effect of an external magnetic field on the radiation pattern.  A magnetic field would change the spin waves in the nanomagnets and that should alter the radiation pattern.  Unfortunately, we could not test this because there are two hindrances. The first is of a practical nature; our anechoic chamber is not equipped to accommodate a magnetic field. Our sample holder cannot be placed within the pole pieces of an electromagnet and even if that were possible, the electromagnetic emission cannot travel through the body of the electromagnet to reach the detecting horn antenna. For that, we will need a ``transparent'' electromagnet which is not available. The second hindrance is more fundamental. It is well known that the magnetic field will affect electromagnetic waves regardless of their origin. Hence we could never determine unambiguously whether any observed change in the radiation pattern is due to a change in the spin wave patterns or merely a manifestation of the effect of an ambient magnetic field on electromagnetic radiation. 

\section{Conclusion}
Topological insulators (TI) sport exotic properties such as spin-momentum locking. While much effort has been expended on unraveling their physics and exploring the subtle material nuances, there is still {\it not a single product in the market based on them} after more than a decade of research by numerous groups. Transforming from the realm of ``science'' to the realm of ``engineering'', here we have demonstrated a novel spintronic  micro-antenna activated by spin injection (with alternating spin polarization) from a three-dimensional topological insulator (3D-TI) into an array of nanomagnets deposited on the surface of the 3D-TI. This may be the first pathway to a marketable product. 

The beam steering is enabled by spin-momentum locking which is a quantum property of topological insulators. In that sense, it is truly a {\it TI-based device}. To date, TIs have been used almost exclusively to inject spin into a ferromagnet to flip its magnetization, which has applications in memory \cite{memory} and perhaps logic \cite{logic}. Those are digital applications. Here, we have demonstrated an analog application along with the unique feature of beam steering. Normally beam steering requires a phased array consisting of multiple antennas, each much larger than the electromagnetic wavelength. Ours is an ultra-miniaturized beam steerer (single element much smaller than the wavelength) which makes it a disruptive technology.

The micro-antenna that we have demonstrated is orders of magnitude smaller than the radiated free-space electromagnetic wavelength and yet it emits efficiently because the activation mechanism is unconventional. Similar antennas (none of them based on a TI) have recently attracted significant attention \cite{raisa,carman,justine,saibal}. This work can open an avenue for the application of topological insulators beyond just memory and logic -- for antennas which are analog devices, and by extension, other analog and communication devices.

\section*{Methods and Materials}
{\it Nanomagnet Array Fabrication}: 
The TI film grown on a PMN-PT substrate is first cleaned in acetone and isopropyl alcohol and the Al electrodes for injecting current are delineated using optical lithography. After delineation of the electrodes, the substrate is spin-coated (spinning rate $\approx$2500 rpm) with
a single layer polymethyl methacrylate (PMMA) resist and subsequently
baked at 110$^{\circ}$C for 2 minutes. Next, electron beam lithography is performed
using a Raith Voyager Electron Beam Lithography system having accelerating voltage of 50 kV and beam current of 300 pA to open windows
for deposition of the nanomagnets. The resists are subsequently developed in
methyl isobutyl ketone and isopropyl alcohol (MIBK-IPA, 1 : 3) for 60 s, followed by a cold isopropyl alcohol (IPA) rinse. A 5 nm-thick
Ti adhesion layer is deposited on the patterned substrate using electron
beam evaporation base pressure 2.3$\times$10$^{-7}$ Torr, followed by the electron
beam deposition of 6 nm-thick Co. The lift-off is carried out by remover
PG solution (a proprietary solvent stripper).

{\it Antenna Measurements}: Measurements of antenna radiation patterns are carried out in an
AMS-8701 Anechoic Chamber, Antenna Measurement System using a 3164-10 Open Boundary Quad-ridged Horn
Antenna. The sample (antenna) is always placed at a distance of 284.5 cm from the horn antenna which ensures that
we are measuring the far-field radiation pattern at all frequencies.

\section*{Supplementary Material}

Supplementary Material ...

\section*{Acknowledgments}
We acknowledge Mr. Michael Suche and Prof. Erdem Topsakal of the Department of Electrical and Computer Engineering at Virginia Commonwealth University for help with microwave measurements in the anechoic chamber. 




\end{document}